# Controlling and Phase-Locking a THz Quantum Cascade Laser Frequency Comb by Small Optical Frequency Tuning

L. Consolino,[1,2,*] A. Campa,[1] M. De Regis,[1] F. Cappelli,[1] G. Scalari,[3] J. Faist,[3] M. Beck,[3] M. Roesch,[3] S. Bartalini[1,4] and P. De Natale[1]

[1]CNR- Istituto Nazionale di Ottica e LENS, Via Nello Carrara 1, I-50019 Sesto Fiorentino, Italy
[2]Dipartimento di Fisica e Astronomia, Università degli Studi di Firenze, I-50019 Sesto Fiorentino, Italy
[3]Institute for Quantum Electronics, ETH Zurich, 8093 Zurich, Switzerland
[4]ppqSense Srl, Via Gattinella 20, I-50013 Campi Bisenzio FI, Italy
*luigi.consolino@ino.it

**Abstract:** Full phase control of THz emitting quantum cascade laser (QCL) combs has recently been demonstrated, opening new perspectives for even the most demanding applications. In this framework, simplifying the set-ups for control of these devices will help to accelerate their spreading in many fields. We report a new way to control the emission frequencies of a THz QCL comb by small optical frequency tuning (SOFT), using a very simple experimental setup, exploiting the incoherent emission of an ordinary white light emitting diode. The slightly perturbative regime accessible in these condition allows tweaking the complex refractive index of the semiconductor without destabilizing the broadband laser gain. The SOFT actuator is characterized and compared to another actuator, the QCL driving current. The suitability of this additional degree of freedom for frequency and phase stabilization of a THz QCL comb is shown and perspectives are discussed.

# 1. Introduction

In the field of frequency metrology a paradigmatic shift has occurred with the invention of optical Frequency Combs [1–3] (FCs, commonly fs-pulse mode-locked lasers). These provide a frequency ruler for any laser emitting within their broad spectral range [4], and can enhance the performances of environmental sensing setups [5], thanks to the combination of the broad spectral coverage with the selectivity and sensitivity of laser-based spectrometers. In fact, a FC is a superposition of a series of monochromatic electromagnetic waves (the FC modes, that can be fully controlled with two global parameters, i.e. repetition rate and offset frequency), coherently spreading along a broad spectral range.

This technology is particularly important in the mid-infrared and terahertz (THz) regions, where many chemical species have characteristic absorption transitions, and where sensing and high-precision spectroscopy on molecular compounds have gained great scientific importance. In particular, a different approach to broadband THz frequency metrology relies on the recent extension of FCs to the far infrared domain. In fact, in the THz region, the generation of pulses by optical rectification of fs-lasers dates back 15 years [6], and already contained all the constituents of a zero-offset FC. However, the comb nature of such emission has been unveiled only several years later [7], and its application to high-precision spectroscopy was later on reported as frequency reference [8]. With the aim of replicating the operating principles of a FC in a very compact solid-state THz source, great efforts have been devoted to study and pursue mode-locked emission regime in THz quantum cascade laser frequency combs (QCL-combs) [9,10].

Indeed, the development of active regions with engineered optical dispersion allows for the simultaneous emission on a large number of equally-spaced longitudinal modes emitted by this kind of devices. Moreover, high third-order nonlinearities in the QCLs active region enable very fast four-wave mixing (FWM) processes that induce mode locking of the emission modes, with a high level of coherence [11]. THz QCL-comb coherence has been deeply analyzed with *Shifted Wave Interference Fourier-Transform* (SWIFT) spectroscopy [12] and *Fourier-transform Analysis of Comb Emission* (FACE) technique [13], proving the high degree of coherence of these sources, as well as their long term stability. Very recently, full phase-stabilization of a QCL-comb at a metrological level has been demonstrated [14], proving that this technology is suited for challenging metrological grade applications. In this latest report, the spacing among the modes and offset frequency of the QCL-comb are stabilized through only one actuator, the device driving current, by exploring two very different frequency ranges. Finally, free running THz QCL-combs have been used as laser sources for dual comb spectrometers [15,16], already providing state-of-the-art resolution in the determination of transitions line centers, when properly referenced to a primary frequency standard [17].

THz QCLs frequency tuning is usually obtained by driving current or heat-sink temperature control, but, recently, a new approach has been followed for single-mode devices. Initially demonstrated for mid-infrared devices [18,19], it has shown how a focused near-infrared excitation of the semiconductor substrate just below the THz QCL active medium allows a frequency coverage exceeding ten times the tuning by current (usually around 4 GHz) [20]. This laser induce frequency tuning (LIFT) mechanism relies on the generation of a non-thermal electron-hole plasma in the vicinity of the back facet of edge-emitting single-plasmon waveguide devices, which changes the optical cavity length and consequently the mode frequency. Application to high-resolution molecular spectroscopy has also been demonstrated [20], and, more importantly, LIFT has also allowed simultaneous frequency and power stabilization of a single mode THz QCL [21].

In this work, we aim to apply the advantages of tuning through electromagnetic radiation to a THz QCL-comb, showing how light influences the device multi-mode emission. In particular, we focus on QCLs based on double-metal waveguides. For these devices there is no optical mode extending in the semiconductor substrate, and we therefore directly excite the laser active region. This condition was discarded in previous works, due to instabilities induced in the QCL emission by the high intensity of the near-infrared laser used for tuning [20]. For this reason, instead of employing intense and focused laser radiation, we use a broadband white light emitting diode (LED) placed outside but close to the cryostat window, realizing a simple, cost effective, robust and low power consumption setup. Clearly, the amount of light coupled to the QCL facet will be much smaller than the focused laser case. For this reason, we are able to operate in a slightly perturbative regime, affecting the laser gain, but not destabilizing it. This new control technique will be called small optical frequency tuning (SOFT), in order to differentiate it from the huge tuning ranges allowed by LIFT. Nevertheless, here attention is rather posed on the possibility of exploiting these small tunings for exact frequency and phase control. In fact, our data and analysis confirm the possibility of using SOFT as external actuator on the QCL-comb degrees of freedom (mode spacing and frequency offset), and we show that SOFT can actually induce full phase control of one of these parameters, phase-locking it to the primary frequency standard. Finally, we study the similarities of this new actuator with respect to driving current, in order to envision a full frequency comb phase locking.

## 2. Experimental Setup

The experimental setup realized for this work is schematically sketched in fig. 1a. The white LED used in this work for the QCL-comb SOFT is mounted in front of the cryostat quartz window, and it is voltage driven by a Agilent 33250A function generator, that delivers the voltage bias and allows modulation of the device.

A detailed description of the QCL-comb device used in this work can be found in [13], it is a heterogeneous cascade laser device emitting in the 2.5–3.1 THz spectral region. It is based on a double-metal ridge waveguide, defined via dry etching, of length 2 mm and width 60 μm, with lateral side absorbers for enhanced transverse mode suppression [22,23]. It is mounted on the cold finger of a helium flux cryostat, and it is driven in continuous wave (CW) mode by a low noise current driver (ppqSense QubeCL-P05). A fast bias tee (Marki Microwave BT-0024SMG) is mounted inside the QCL cryostat, as close as possible to the laser device, and allows electrical retrieval of the QCL-comb inter-mode beatnote frequency ($f_{IBN}$), corresponding to the spacing among the modes of the QCL-comb. In the current range 305–335 mA, the device displays a single IBN signal at around $f_{IBN}$ = 19.8 GHz, with a spectral width in the order of few tens of kHz.

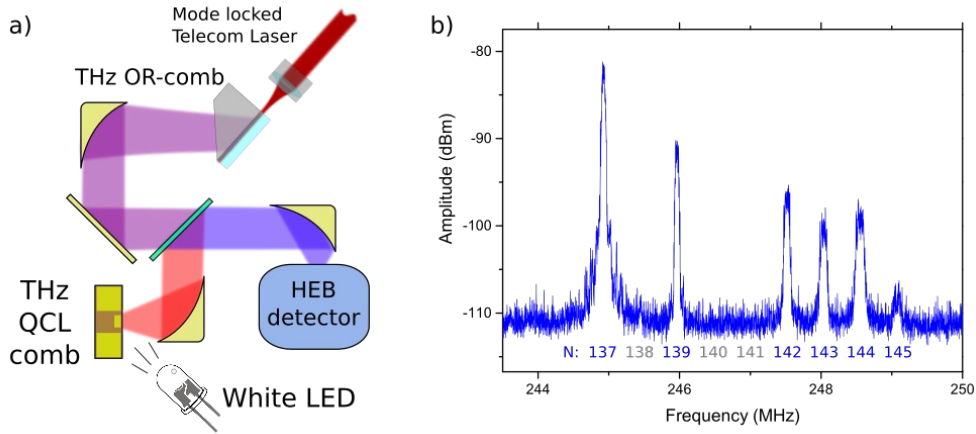

Fig 1: a) Sketch of the experimental setup. The OR-comb and QCL-comb are spatially superimposed and mixed on a hot-electron bolometer (HEB) detector. At the same time, a white LED placed outside the QCL cryostat window allows SOFT of the device. b) Sample spectrum of the down-converted radio frequency optical beatnotes (OBNs) acquired at around 246 MHz. 6 most intense OBNs are distinguishable, and their mode order has been retrieved by dividing the absolute THz frequencies by their frequency spacing ($f_{IBN}$).

The QCL-comb emission is superimposed to the radiation of a fully-stabilized free-standing THz frequency comb, generated via optical rectification (OR-comb) in a lithium niobate single mode waveguide [7]. The absolute frequencies of the modes of the OR-comb are known with a $10^{-13}$ accuracy (depending on the frequency standard employed in our laboratory, i.e. a GPS-Rb-Quartz frequency chain), and are fixed by tight phase stabilization of the pumping telecom femtosecond laser repetition rate (Menlo Systems FC1500). The two beams are mixed onto a fast hot electron bolometer detector (HEB - Scontel RS0.3-3T1). The resulting multi-heterodyne detection, with a close-to-integer Vernier ratio, allows down conversion of the QCL-comb emission into a series of equally spaced, optically detected beatnotes (OBNs) in the radio frequency domain (RF-comb). These OBNs carry full information on the parenting modes and, since the OR-comb modes frequencies are known, they can be used to retrieve the absolute frequencies of the QCL-comb modes, with the technique thoroughly described in [24]. In this way, thanks to the absolute frequencies retrieval, by dividing them by the intermode beating frequency and rounding up to the nearest integer, the QCL-comb mode order has been extracted, and labeled in fig. 1b.

Moreover, the OBNs allow monitoring the QCL-comb modes frequencies, tracking their evolution in time. To this purpose, the acquisition of the OBN frequencies is performed by means of a real-time spectrum analyzer (Tektronix RSA5106A). In the operating conditions used for this work (QCL heat-sink temperature ~20 K and driving current in the range 305–335 mA), the spectrum of the QCL shows 11 intense modes, while the other emitting modes have an intensity more than 10 times lower [13]. In the analysis presented in this work we have preferred faster yet less sensitive acquisition scans, and therefore only the 6 most intense modes can be retrieved and will be considered. A sample acquisition of the RF-comb spectrum in these conditions is shown in fig. 1b.

## 3. Results and Discussion
### SOFT Characterization

The QCL-comb emitted modes have frequencies that can be parametrized as:

$$f_N = N \cdot f_{IBN} + f_{off} \quad \text{eq. 1}$$

where $N$ is the order of the mode considered and $f_{off}$ is the QCL-comb frequency offset. By simply deriving this equation, it is possible to retrieve the contribution of IBN and offset tuning on the emitted comb frequencies. Moreover, by keeping the OR-comb modes' frequencies fixed, these variations are transferred to the acquired RF-comb modes:

$$\frac{\Delta f_N}{\Delta k} = \frac{\Delta f_{OBN}}{\Delta k} = \pm N \frac{\Delta f_{IBN}}{\Delta k} \pm \frac{\Delta f_{off}}{\Delta k} \quad \text{eq. 2}$$

where $k$ is one tuning parameter, i.e. driving current, heat-sink temperature or LED light intensity. The sign uncertainties are lifted once the directions of the IBN and offset tunings are retrieved, knowing that they can depend on the actuator used.

In fact, thanks to the developed experimental setup, down-converting the IBN to RF frequencies by mixing it with a proper synthesized frequency, it is possible to acquire both the IBN and the OBNs simultaneously on the same spectrum analyzer, and trace their frequency shifts while tuning the device parameters. Therefore, using equation 2, it is possible to discriminate the spacing and offset tunings for all actuators. Table 1 summarizes how the two comb parameters drift with respect to driving current, heat-sink temperature, and LED light intensity. It is interesting to note that the signs of temperature and LED tunings are not compatible with each other, confirming that the effect of the SOFT is not merely a thermal effect. Moreover, quite obviously, the temperature tuning is a slow mechanism, that is neither suited for a fast control of the comb parameters, nor for a tight phase control. As a consequence, in the following we will focus the attention on current and LED tunings, and, in order to study the effect of a modulation applied to these two actuators, we acquire the time evolution of the intermodal frequency shift as well as of the optical beatnotes.

| **Increment of** | **IBN frequency** | **Offset frequency** |
|:---:|:---:|:---:|
| Driving current | + | - |
| Heat-sink temperature | - | - |
| LED light intensity | - | + |

Table 1: Frequency tuning shifting direction for an increment of the three actuators: driving current, heat-sink temperature and light emitting diode (LED) light intensity.

Fig. 2 shows the effect of applying a modulation on the QCL-comb through SOFT. The LED is biased with a 8 V voltage, and is modulated sinusoidally with a 1 V amplitude at 30 Hz. The time evolution of the six, simultaneously acquired, OBNs and IBN is reported in fig. 2a. From these data we can extract an IBN frequency shift ($\Delta f_{IBN}$) of 107.0(3) kHz, leading to a tuning coefficient of -107.0(3) kHz/V. Moreover, the individual OBN frequency shifts ($\Delta f_{OBN}$) are retrieved, and their overall dependence on the mode order is shown in fig. 2b. This linear dependence is the result of the modulation on the IBN and propagating along the comb modes. Combining eq. 2 with the tuning directions reported in table 1, we can calculate the offset frequency shift ($\Delta f_{off}$) corresponding to the 1 V LED bias modulation, leading to a mean value of 15.98(5) MHz/V. All the details of these analysis are reported in supplementary materials section.

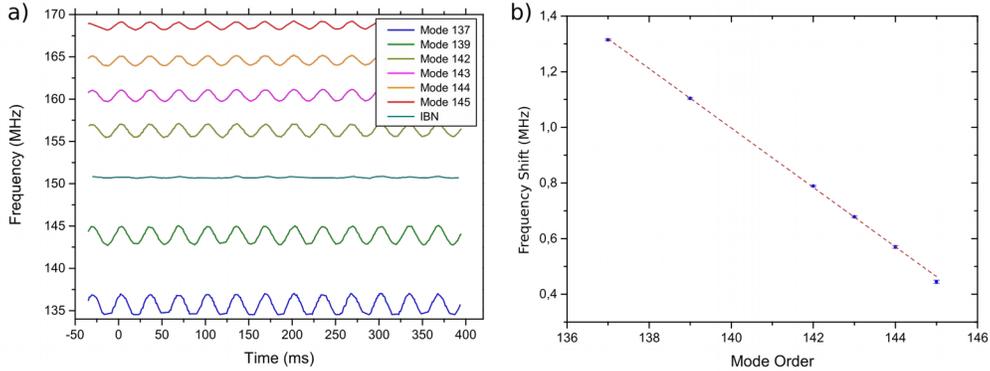

Fig. 2: SOFT modulation. a) The time evolution of the optical beatnotes (OBNs) and intermode beatnote (IBN) is recorded while the device is SOFT modulated at 30 Hz. The frequency shift of each OBN is measured and reported in panel b). The linear dependence on the mode order is due to IBN frequency modulation spreading along the comb modes.

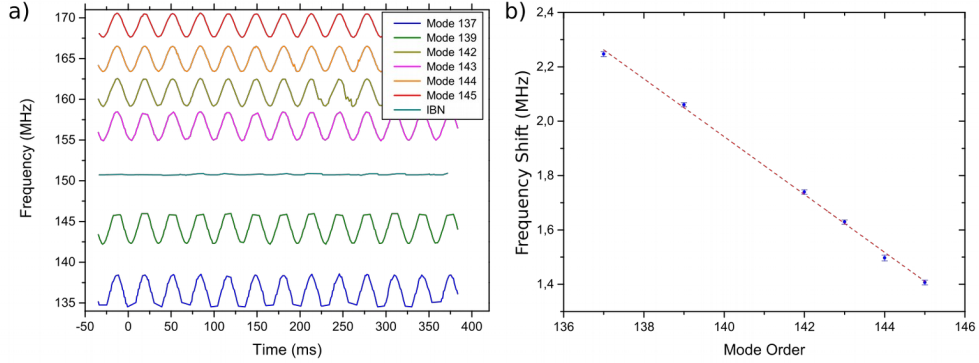

Fig. 3: Driving current modulation. a) The time evolution of the optical beatnotes (OBNs) and intermode beatnote (IBN) is recorded while the device current is modulated at 30 Hz. The width of each OBN is measured and reported in panel b). The linear dependence on the mode order is due to IBN frequency modulation spreading along the comb modes.

The same procedure is applied to the device driving current modulation, and the results are shown in fig. 3. The device is driven at 320 mA, and the amplitude of the modulation (50 μA at 30 Hz) is chosen to result in an IBN amplitude modulation close to the previous case. For the current modulation an IBN shift of 110.3(3) kHz leads to a tuning coefficient of 2.206(6) MHz/mA. Figure 3a shows the OBNs and IBN time evolution, while the OBNs frequency shift amplitude linear dependence on the mode order is shown in fig. 3b. The offset modulation coefficient is retrieved as in the previous case, leading to -347.8(2.2) MHz/mA, while all the details of the analysis are reported in supplementary materials section.

| Actuator | $\Delta f_{off}/\Delta k$ | $\Delta f_{IBN}/\Delta k$ | $\Delta f_{off}/\Delta f_{IBN}$ |
|---|---|---|---|
| SOFT | +15.98(5) MHz/V | -0.1070(3) MHz/V | -149.3 |
| Driving current | -347.8(2.2) MHz/mA | +2.206(6) MHz/mA | -157.7 |

Table 2: Frequency offset and intermode beatnote frequency tuning coefficients for both SOFT and driving current actuator. The ratio between the two coefficients is a good parameter to compare the tuning characteristics of the two actuators on the device.

All the previous results, related to SOFT and current tunings have been summarized in table 2. The driving current tuning coefficients are reported, while the *moduli* of the two SOFT tuning coefficients depend also on other non-device-related parameters, such as the LED optical power effectively impinging on the QCL (depending mainly on the distance between the LED and the QCL, and the cryostat window transmittivity). For this reason, in order to compare the actuators, we can calculate the ratio between the two coefficients, shown in the last column of table 2. It is clear that, while the two actuators have indeed similar effect on the QCL-comb, they do not act exactly in the same way.

The LED illumination creates an electron-hole plasma in a thin slice (1-2 μm) close to the laser cavity edges. Depending on the intensity, the amount of the photogenerated electrons will change more or less significantly the complex refractive index of the semiconductor. From the experimental geometry and the laser dimension we can estimate that the power illuminating the laser edge is $10^{-4}$ of the LED power, that is around 13 mW overall. From the LED power coupled to the laser (about 1.5 μW) we would expect then a tuning of the laser lines $10^{-5}$ times lower than what observed for the laser case of Ref. [20]. As already stated above, in the case of our experiment the refractive index change is happening in the active region of the laser where the electrons are confined in the multiquantum well heterostructure and not in the bulk semiconductor (as in Ref. [20]).

The broadband nature of the LED illumination leads to a non-selective injection of carriers in the active region populating also higher levels in the subbands. The photoexcited electrons will relax via intraband scattering in the lower subbands changing the effective temperature distribution of the electrons in the lasing levels (see also Ref. [19] for a similar discussion) and locally perturbing the gain. This is different with respect to the selective carrier population change achieved by varying the current and the population of the upper lasing state through proper resonant tunneling injection. So the two actuators impact differently the electron distribution and dynamics in the laser active region.

In order to confirm and quantify this difference, two tests have been performed. First, we can measure the two different fixed points for the two actuators. A frequency comb fixed point is the only frequency that remains stationary while the comb expands or contracts around it, due to a frequency tuning. Therefore the fixed points can be extracted from eq. 2 and eq. 1, by simply considering:

$$\frac{\Delta f_{fix}}{\Delta k} = N \frac{\Delta f_{IBN}}{\Delta k} + \frac{\Delta f_{off}}{\Delta k} = 0 \quad \text{eq. 3}$$

Where the sign ambiguity of eq. 2 has been lifted by the measurements.

Finally:

$$f_{fix} = f_{off} - \frac{\Delta f_{off}}{\Delta f_{IBN}} f_{IBN} \quad \text{eq. 4}$$

Dialing in the numbers we calculate $f_{fix} = 2.922$ THz for the SOFT actuator, and $f_{fix} = 3.060$ THz for the driving current. The two fixed points are outside the detected emission bandwidth of the comb, and despite being close (138 GHz), they are clearly distinguished thus allowing, in principle, an independent control of the comb degrees of freedom.

The second test that proves the difference between the two actuators is to stabilize the frequency of one comb degree of freedom by simultaneously applying modulations through the driving current and the LED light, with ad-hoc amplitudes and a fixed phase difference. In fact, if the two actuators were the same, there would be no linear combination of the two modulations minimizing only one degree of freedom. Fig. 4 shows the effect of the two modulations at 30 Hz, employed to stabilize the frequency of the intermode beatnote, by finely tuning the ratio of the amplitudes and the relative phase. As in the previous cases, fig. 4a shows the time evolution of the OBNs but, in this case, the width of their frequency shifts is constant throughout the detected comb emission (fig. 4b). This effect is due to negligible contribution of the stabilized $f_{IBN}$ parameter on the width of the OBNs.

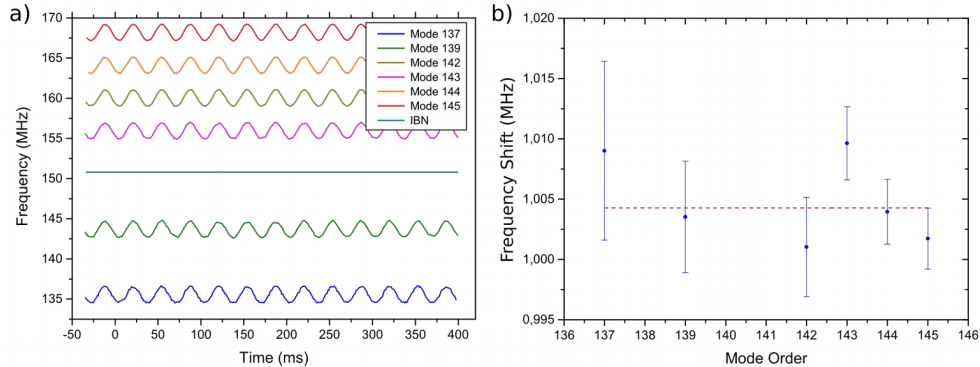

Fig. 4: SOFT and Driving current modulation. a) The time evolution of the optical beatnotes (OBNs) is recorded while the SOFT and driving current modulations are simultaneously applied onto the device. The amplitudes and relative phases are tuned in order to minimize the intermode beatnote (IBN) width. b) Width of each retrieved OBN. The absence of dependence on the mode order is due to negligible IBN frequency modulation.

**IBN Phase control through SOFT**

One intriguing possibility is to use the SOFT actuator to obtain full phase control on the QCL-comb emission, by controlling one comb degree of freedom with SOFT and the other with the driving current. With respect to the driving current, the LED actuator is expected to be a smaller bandwidth control, and, as a consequence, it has been chosen to phase lock the comb mode spacing, showing a characteristic width in the order of few tens of kHz. To this purpose, the IBN signal is isolated, down converted to RF frequencies, and compared to a local oscillator on a standard phase detector unit. The phase detector output signal is properly elaborated by a proportional-integral-derivative (PID) controller, and fed back to the LED driving unit, closing a simple feedback loop acting on the SOFT actuator. Once all the parameters are optimized, we achieve a stable phase-locking loop (PLL) on the IBN signal. Fig. 5 compares the IBN free-running spectrum (fig. 5a), the SOFT phase-locked one (fig. 5b) and the same IBN, phase-locked through driving current (fig. 5c). The SOFT PLL (PLL1) has an electronic bandwidth of 15 kHz but, clearly, the locking figure-of-merit is not as good as in the driving current case. The effect of this non-perfect phase lock can be evidenced by applying a modulation on the device driving current, while leaving the IBN free-running or phase-locked with through SOFT. This comparison is shown in fig. 5d. While it is clear that the activation of the SOFT PLL1 reduces the effect of the current modulation, the OBNs frequency shifts show a residual linear dependence on the comb mode order (slope 0.0045±0.0018), due to a non-perfect locking loop.

In order to attempt complete phase control of the comb emission, we activate the PLL on the comb mode spacing through SOFT and simultaneously stabilize the comb offset. Following the work done in [14], this latter stabilization is pursued by phase locking one selected OBN to the optically rectified THz comb through the device driving current (PLL2). Fig. 6 show the two PLL error signals, simultaneously acquired. Unfortunately, the two locking loops cannot operate simultaneously, in fact, every time PLL2 locks one selected OBN, PLL1 is interrupted, and starts working again when PLL2 is interrupted. The reasons for the mutual exclusivity of the two loops is the non-orthogonality of the two actuators. In fact, every time one of the two loops is active, the phase noise affecting the locked degree of freedom is transferred to the free-running one, broadening the relative beatnote signal, and disturbing the locking loop.

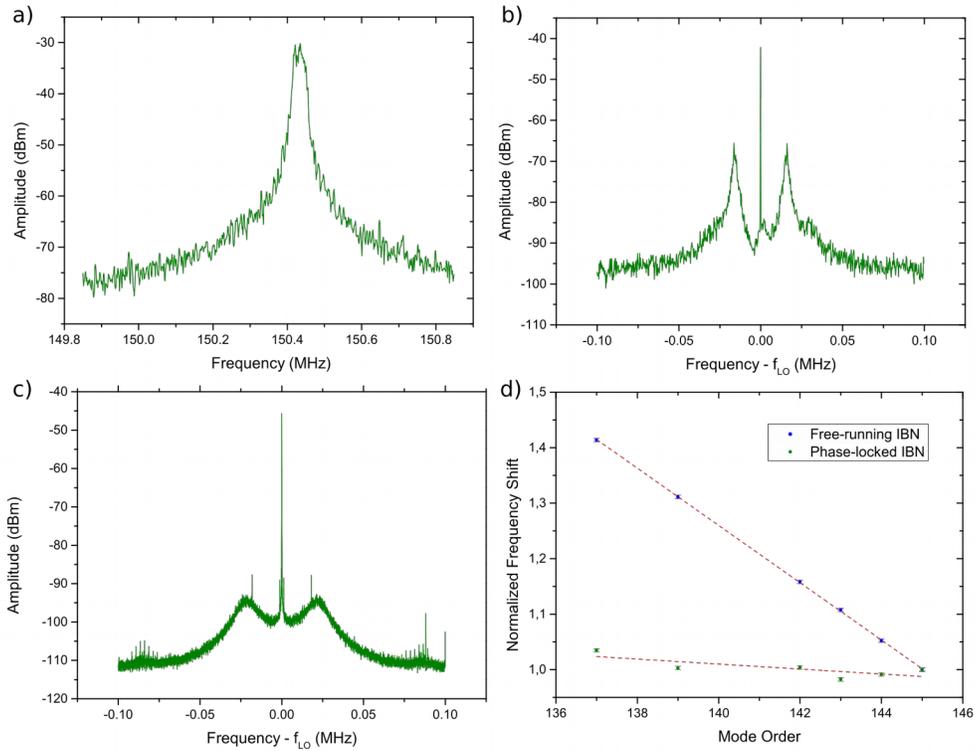

Fig. 5: Intermode beatnote (IBN) phase locking. a) the free-running spectrum of the IBN, down-converted to RF frequencies and acquired with a resolution bandwidth (RBW) of 300 Hz. b) IBN phase-locked using the SOFT actuator to a local oscillator at frequency $f_{LO}$ =150 MHz. c) IBN phase locked through driving current actuator. Both b) and c) have been acquired at RBW = 300 Hz. d) Normalized frequency shifts of the optical beatnotes (OBNs) while applying a current modulation and with a free running (blue) and phase locked IBN through SOFT (olive).

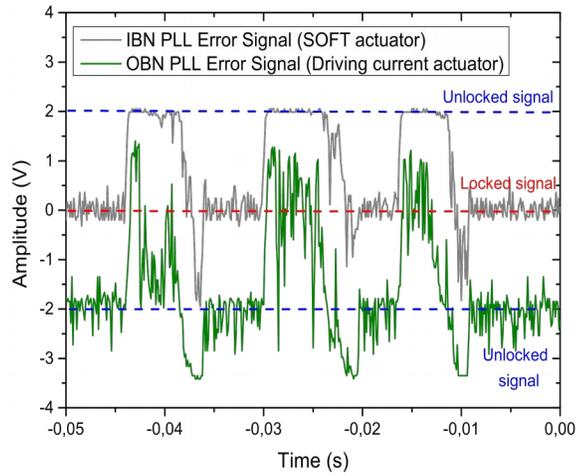

Fig. 6: The time evolution of phase-locking loop (PLL) on the itermode beatnote (IBN) through SOFT (PLL1 - grey) and PLL of one optical beatnote (OBN) through driving current (PPL2 – green) signals are shown. The attempts to simultaneously run the two loops fail, due to the non-orthogonality of the two used actuators.

## 4. Conclusions

In this work we have demonstrated how LED illumination can be used to tune the emission of a THz QCL-comb, entering a new regime of control with small optical frequency tuning (SOFT). We have reported on how the SOFT actuator compares with the device driving current, showing that the two actuators are similar but distinguishable. Finally, tight phase locking of the QCL IBN to a primary frequency standard is demonstrated, thus enabling phase control on the free spectral range of the THz QCL-comb emitter. It has to be noted that ambient laboratory light can have an impact on the laser emission frequencies and also on their stability. For this reason all the measurements reported in this paper were performed in the dark. In cases where light perturbation can affect the device performances particular filters in the cryostat window can be adopted, in order to filter out the ambiet light.

The additional SOFT actuator, with respect to current, temperature and RF injection locking, significantly widens perspectives of full control of THz comb emission in an increasing number of scenarios. In particular, for future wireless communication networks, such LED control of QCLs can enable practical Optical-to-THz multichannel encoding of information, recently demonstrated for single THz frequencies by photomixing telecom frequencies [25]. Indeed, THz carrier frequencies can carry more information than microwave range carriers, while propagation in scattering atmosphere is much more favourable than telecom wavelengths. Of course, in-field deployment of such optical-to-THz converters will depend on reliable close to room temperature operation of THz QCL-combs [28]. Moreover, similarly to classical encoding of information, transduction of quantum information could be envisaged from telecom/visible lasers to THz QCL combs once, for these latter, squeezing/entanglement properties will be demonstrated [in preparation].

**Acknowledgements:** The authors acknowledge financial support from ERC grant CHIC (724344) the Qombs Project, EU-H2020-FET Flagship on Quantum Technologies grant no. 820419, the Italian ESFRI Roadmap (Extreme Light Infrastructure-ELI), EC–H2020 Laserlab-Europe grant agreement 654148, the EC Project 665158 (ULTRAQCL).

**Competing interests:** The authors declare no competing interests.

**Data availability:** The data supporting the findings of this study are available from the corresponding author upon reasonable request.